\documentclass{rsos}
\usepackage[numbers]{natbib} 
\begin{document}
\title{The Dispersal of Planet-forming discs: Theory confronts Observations}
\newcommand{\angstrom}{\mbox{\normalfont\AA}}
\newcommand{\micron}{$\mu$m}
\def\mnras{{MNRAS}}
\def\apj{{ApJ}} 
\def\apjl{{ApJL}}
\def\aj{{AJ}} 
\def\aap{{A\&A}}
\def\pasa{{PASA}}
\def\nat{{Nature}}
\def\pasp{{PASP}}
\def\araa{{ARA\&A}}
\def\ssr{{Space Science Reviews}}
\def\apjs{{ApJS}}
\def\icarus{{Icarus}}

\author{Barbara Ercolano$^{1,2}$, Ilaria Pascucci$^{3,4}$}
\address{$^1$Universit\"ats-Sternwarte M\"unchen, Scheinerstr. 1, 81679 M\"unchen, Germany\\
$^2$Excellence Cluster Origin and Structure of the Universe,
Boltzmannstr.2, 85748 Garching bei M\"unchen, Germany\\
$^3$Lunar and Planetary Laboratory, The University of Arizona, Tucson, AZ 85721, USA\\
$^4$Earths in Other Solar Systems Team, NASA Nexus for Exoplanet System Science\\}
\subject{Astrophysics}
\corres{ercolano\@usm.lmu.de}
\begin{abstract}
Discs of gas and dust around Myr-old stars are a by-product of the star formation process and provide the raw material to form planets. Hence, their evolution and dispersal directly impact what type of planets can form and affect the final architecture of planetary systems. Here, we review empirical constraints on disc evolution and dispersal with special emphasis on transition discs, a subset of discs that appear to be caught in the act of clearing out planet-forming material. Along with observations, we summarize theoretical models that build our physical understanding of how discs evolve and disperse and discuss their significance in the context of the formation and evolution of planetary systems. By confronting theoretical predictions with observations,  we also identify the most promising areas for future progress.
\end{abstract}
\begin{fmtext}
............
\end{fmtext}
\maketitle
\section{Introduction}
Planets form from the dust and gas contained in discs around young stars. These discs are a natural consequence of the star-formation process, implying that all young stars have the potential to host a planetary system.
Circumstellar discs are observed to evolve and finally disperse over a timescale of a few Myr (Section~\ref{sect:obs}), which is comparable to the timescales for planet formation by the core accretion process
\cite{2014prpl.conf..643H} and to migration timescales for giant planets
\cite{2014prpl.conf..667B}. This implies that the processes driving the evolution and dispersal of discs play a crucial role in shaping new planetary systems and likely contribute to the observed exo-planets diversity
\cite{2012MNRAS.422L..82A,2012A&A...541A..97M,2015MNRAS.450.3008E}. 

In the standard picture, the evolution of the surface density of {\it young} circumstellar discs is controlled by viscous accretion while the late evolution and final dispersal by star-driven photoevaporation (see Section~\ref{sect:theory} where we also discuss possible recent revisions). A class of objects that are thought to be in an intermediate state between possessing an optically thick (dust) disc and being disc-less, known as ``transition discs'', may provide us with important insights on the disc dispersal mechanisms, and has been the subject of recent observational \cite{2014prpl.conf..497E} and theoretical \cite{2016PASA...33....5O} reviews.
It is however becoming clear that transition discs are in reality a diverse class of objects. Some of them may not actually be short-lived objects caught in the act of dispersing their discs, but rather produced by a different rarer and longer lived phenomenon, \cite{2016PASA...33....5O}. 



This Review aims at making a stronger connection between theory and observations, by exploring the predicting power and limitations of current disc evolution and dispersal models and by discussing how observations inform us on how discs disperse. In Section~\ref{sect:obs} we summarize empirical constraints on the evolution and dispersal of circumstellar gas and dust while  Section~\ref{sect:theory} is an overview of theories of disc dispersal. Section~\ref{sect:thevsobs} confronts theoretical predictions with observations while Section \ref{sect:implications} explores the implications of disc dispersal processes on planet formation and planetary architectures. We conclude with Section~\ref{sect:summary} by summarizing current theoretical successes and shortcomings and by discussing which investigations are most promising to clarify how circumstellar discs disperse.

\section{Observations of dispersing discs}\label{sect:obs}

Young stars are historically classified either from their broad-band continuum emission at infrared wavelengths (SED), probing circumstellar dust, or from optical spectroscopy, probing surrounding gas. The first classification groups young stars into 4 classes that trace from the formation of the star and a massive disc (Class~0 and I) through the so-called protoplanetary disc phase (Class~II) into the dispersal of the disc (Class~III), e.g.  \cite{1984ApJ...287..610L}. Recent observations hint at the presence of giant planets in Class~II discs \cite{2015Natur.527..342S} and structures in Class~I sources that indicate substantial disc evolution linked to planet formation, e.g. \cite{2015ApJ...808L...3A}. As such, we will not use the historical term {\it protoplanetary discs} for Class~II sources but rather call them {\it planet-forming discs}.
The second classification groups young stars into {\it classical}- (C) or {\it weak}-line (W) T~Tauri stars (TTs) mostly based on the strength of the H$\alpha$ emission line  \cite{1988cels.book.....H}. CTTs are actively accreting disc gas.
It was noted early on that most WTTs are associated with Class~III spectral energy distributions 
\cite{1993prpl.conf..689M}. Thus, understanding how circumstellar gas and dust disperse requires observations that can trace the evolution from Class~II to Class~III and from CTTs to WTTs. As X-ray emission is elevated for all TTs, X-ray surveys of star-forming regions are also an important complement to optical and infrared observations to recover the true pre-main sequence stellar population  \cite{2004ApJ...614..267F}.

\subsection{Disc lifetime vs dispersal time}\label{sect:disc_clearing}
As circumstellar discs form, gas should account for almost all of the disc mass while dust only for about $\sim$1\%. However, most of our understanding of how discs evolve and disperse is based on observations tracing the dust component. 
In addition, planet-forming discs are located at hundreds of parsecs and are just a few hundred AU in size, meaning that most observations do not spatially resolve them. One of the goals of this section is to summarize what we know about disc lifetimes, i.e. how long it takes to completely disperse a Class~II disc, and about dispersal time, i.e. how long it takes for a disc to transition from Class~II to Class~III. We should note upfront that these times are linked to the ages of young  star-forming regions and associations,  which are not known in absolute sense below $\sim$20\,Myr 
\cite{2014prpl.conf..219S}.

\subsubsection{Dust diagnostics}\label{sect:dust}
\begin{figure}[b]
\centering
\includegraphics[width=1 \textwidth]{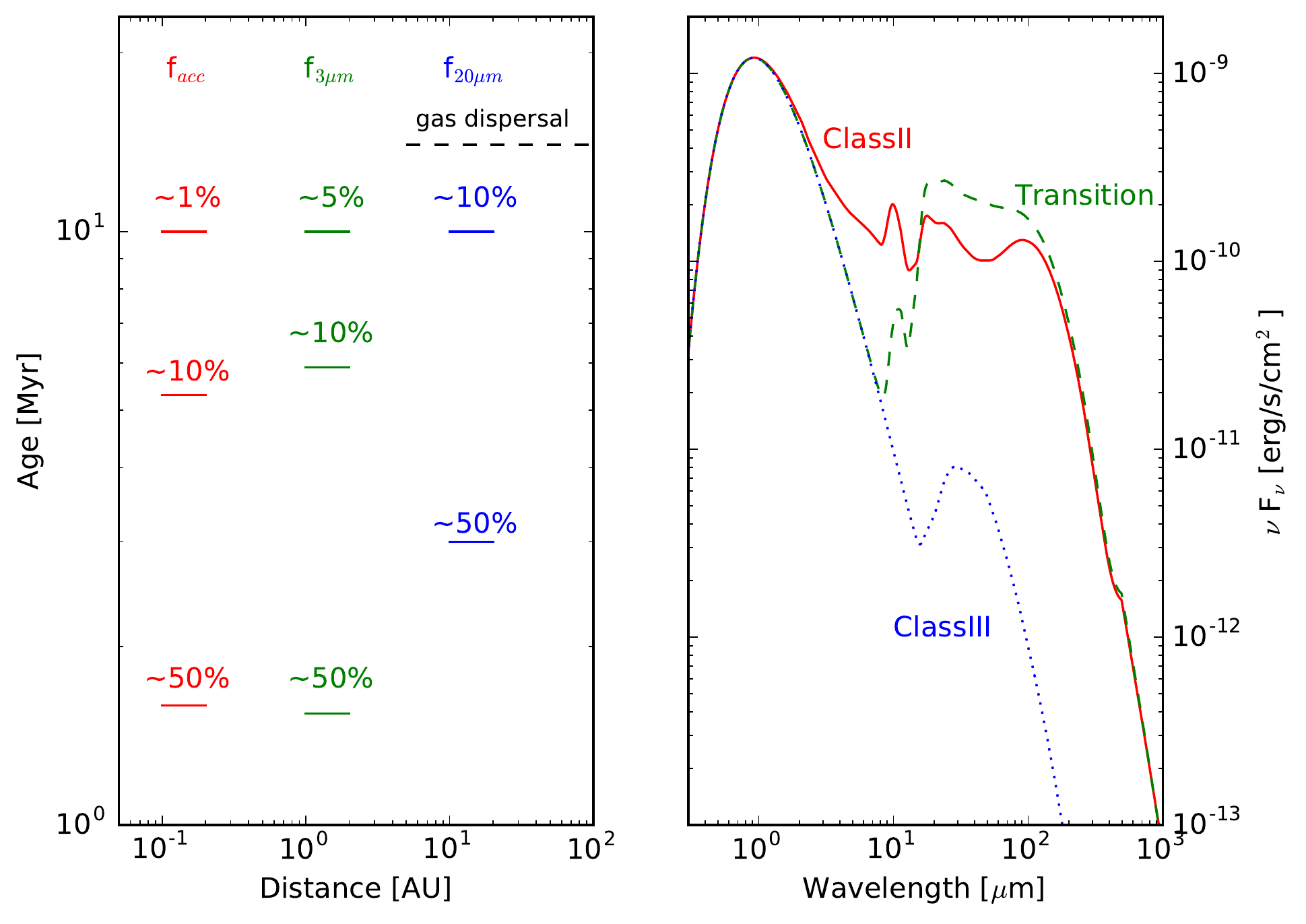}
\caption{Left panel: Figure summarizing the dispersal time of circumstellar discs as a function of tracers (gas and dust) and distance from the star. Right panel: Sample SEDs illustrating a full ClassII disc, a transition disc with a dust gap at 2\,AU, and a tenuous ClassIII debris disc (adapted from \cite{2010pdac.book..263P}).}\label{fig:diskfreq}
\end{figure}

The advent of infrared observatories such as the {\it Spitzer} Space Telescope has enabled sensitive surveys of nearby star-forming regions and associations that revealed the population of young stars with dust discs, i.e. those displaying excess emission above the stellar photosphere at infrared wavelengths. The most recent and comprehensive analysis of these infrared data has been carried out in a recent review \cite{2014A&A...561A..54R} and we refer the reader to their Section~6 for a comparison to previous work.
In short, the fraction of Class~II discs shows a clear decay with time and reaches zero at $\sim 10-20$\,Myr. When assuming that the dust disc fraction decreases exponentially with time, the e-folding time is $\sim$2-3\,Myr at 3-12\,\micron{} and $\sim$4-6\,Myr at 22-24\,\micron{} (see also Figure~\ref{fig:diskfreq} left panel) . {\bf The e-folding time of $\sim$2-3\,Myr is often reported as the disc lifetime.}
While 3-12\,\micron{} wavelengths mostly trace disc material at $\sim$1\,AU around sun-like stars, 22-24\,\micron{} trace dust further away at tens of AU.
 The longer e-folding time at longer wavelengths could either result from inside-out clearing of primordial dust or from second-generation dust produced by collisions of km-size planetesimals that formed in the first few Myr of disc evolution 
 \cite{2008ApJ...672..558C}. It is worth noting that the first scenario implies a  long disc clearing time of a few Myrs. The dust disc lifetime as a function of stellar mass is less well characterized but appears to be shorter for higher- ($\ge 2\,M_\odot$) than lower-mass stars \cite{2006ApJ...651L..49C,2015A&A...576A..52R}. Expanding upon these findings, it was shown that there is an overabundance of discs around lower-mass stars ($\le 0.6\,M_\odot$) in the 5-12\,Myr-old Collinder~69 cluster \cite{2012A&A...547A..80B}, although no difference is found in the spatial distribution and disc fraction in younger clusters \cite{2011MNRAS.416..439E,2013MNRAS.428.3327K}.

Disc fraction estimates are not extended to far-infrared and longer wavelengths as  stellar photospheres red become undetectable at typical distances of hundreds of pc. Nevertheless, even early millimeter surveys could show that the distribution of dust disc masses (hereafter, $M_{\rm dust}$) around solar-mass stars in the young ($\sim$2-Myr) Taurus star-forming region is significantly different than that around 10-30\,Myr-old stars 
\cite{2005AJ....129.1049C}. The sensitivity of ALMA has recently enabled to survey all Class~II discs in the nearby Lupus \cite{2016ApJ...828...46A}, Chamaeleon~I \cite{2016ApJ...831..125P}, and Upper~Sco \cite{2016ApJ...827..142B} regions, thus complementing the previous SMA Taurus millimeter survey \cite{2013ApJ...771..129A}. When adopting the same assumptions to analyze these datasets, it is found that the mean $M_{\rm dust}$ in the 1-3 Myr-old regions of Taurus, Lupus, and Chameleon~I is about 3 times higher than in the 5-10\,Myr-old Upper~Sco association and that the three young regions share the same $M_{\rm dust}-M_*$ scaling relation\footnote{$M_{\rm dust} \propto (M_*)^{1.3-1.9}$ for Chamaeleon~I. The range in the power law index reflects two extremes of the possible relation between the average disc temperature and stellar luminosity} while Upper~Sco has a steeper relation \cite{2016ApJ...831..125P}. This latter finding may result from efficient inward drift of millimeter grains around the lower-mass stars and does not contradict the longer disc lifetimes inferred from infrared observations.  Interestingly, the dust mass of Class~III objects in Upper~Sco, most likely debris or evolved transition disks \cite{2016ApJ...827..142B}, is at least 40 times lower than the average disc mass of 1-3\,Myr-old Class~II objects \cite{2016ApJ...831..125P}. Similarly, discs around higher mass Herbig Ae/Be stars with ages $\ge$10\,Myr, most likely debris discs, are at most 0.5\,M$_{\oplus}$ while younger stars can have discs more than an order of magnitude more massive \cite{2013MNRAS.435.1037P}.
{\bf Overall, observations demonstrate that dust discs evolve significantly from $\sim$1 to 10\,Myr. The fraction of Class~II discs drops from $\sim 70$\% to 10\% and the dust mass of the surviving 10\,Myr-old Class~II discs is reduced by a factor of $\sim 3$ on average.} These observables are consistent with the fast evolution and dispersal of dust in the protosolar nebula as inferred from studies of meteorites and from constraints on the formation time of asteroids, planets, and moons in the Solar System (e.g. Figure 9.5 in \cite{2010pdac.book..263P}). 

The  dust clearing time, the time for a source to transition from Class~II to III, is measured via multi-wavelength infrared observations.
The classical definition of {\it transition} discs goes back to early work \cite{1989AJ.....97.1451S} which included in this group young stars with significantly reduced or no near-infrared excess but large mid- to far-infrared excess emission, see Figure~\ref{fig:diskfreq} right panel for comparison SEDs. This SED type hints to the development of an inner dust cavity that may represent the initial step of an inside-out disc clearing process. While this approach has clear limitations (e.g. in the hole size that can be detected and its sensitivity to small dust grains \cite{2009ApJ...703.1964F, 2014prpl.conf..497E}), SEDs remain the most used tool to identify discs in different evolutionary stages. Based on SEDs, it was realized early on that the fraction of these {\it transition} discs is small, of order 10\% of the Class~II discs 
\cite{1990AJ.....99.1187S} implying a clearing time of a  few $10^5$ years, much shorter than the disc lifetime. It has recently been reported that the  fraction of transition discs increases from 8\% at $\leq$3Myr to about 45\% at $\sim$10\,Myr \cite{2016ApJ...832...87B}.
As the transition disc fraction is defined as the number of transition over the total number of discs and the overall disc fraction decreases by a factor of $\sim$5 in regions older than $\sim$10\,Myr, the above 
results are expected for a time-independent dust clearing time. The existence of another type of transition discs, so-called
{\it homologously depleted} discs has been proposed \cite{2009ApJ...698....1C}. This definition has been applied to disks that have overall weaker infrared excess emission than Taurus discs but stronger emission than debris discs (see also \cite{2010ApJ...708.1107M}). This group would include discs whose surface density decreases homogeneously at all radii with time thus representing a different path of clearing. The disc clearing time, when combining {\it transition} and {\it homologously depleted} discs, is $\sim1$\,Myr 
\cite{2009AJ....138..703C}, a significant fraction of the disc lifetime.
However, it has been successively shown by means of detailed radiative transfer modeling that most of the {\it homologously depleted} discs are consistent with optically thick discs where dust has settled to the disc midplane rather than discs transitioning from Class~II to III through a radial homogeneous draining \cite{2011MNRAS.410..671E}. A re-evaluation of the SEDs of over 1,500 sources in 15 nearby star-forming regions, now considering settled dust discs, finds that less than 2\% of the discs lie in the homogenous draining regime  \cite{2013MNRAS.428.3327K}. This result, combined with the aforementioned observations, suggests that {\bf most discs disperse via inside-out clearing and the disc clearing time is indeed short, a few $10^5$ years}.

Recently, a comprehensive review of the observational properties of {\it transition} discs has been presented  \cite{2014prpl.conf..497E}. What transpires is that the SED-classified {\it transition} discs are a heterogeneous group of objects. While several of the SED-inferred dust cavities have been now confirmed via millimeter imagery 
\cite{2009ApJ...704..496B,2011ApJ...732...42A,2013ApJ...775..136R},  some cavities are also seen in polarized near-infrared light 
(e.g. RX~J1604-2130, \cite{2012ApJ...760L..26M}) while others are not (e.g. SR~21, \cite{2013ApJ...767...10F}). About half of the {\it transition} discs have high millimeter fluxes indicative of large disc masses ($M_{\rm disc}>5$\,M$_{\rm J}$) while the other half are weak millimeter sources 
\cite{2016A&A...592A.126V}. Their mass accretion rates (discussed in the next section) can be as high as that of full discs 
\cite{2014A&A...568A..18M} or much lower 
\cite{2010ApJ...712..925C}. 
This diversity points to different physical mechanisms responsible for a transition-like SED. Since some of these mechanisms may not lead to disc dispersal, the disc clearing time of a few $10^5$ years is an upper limit and we will discuss in Section~\ref{sect:thevsobs} ways to identify those discs that are truly in transition.
We caution that the large number of dust-depleted cavities at mm wavelengths (1 in 3 for the mm-bright discs \cite{2011ApJ...732...42A}) should not be used to estimate the disc clearing time since such cavities can be induced by gas pressure bumps at snowlines or by forming planets 
\cite{2012ApJ...755....6Z}, physical mechanisms that are not truly dispersing disc material (see Section~\ref{sect:theory}).

\subsubsection{Gas diagnostics}\label{sect:gas}
Similarly to the infrared surveys, the gas lifetime can be obtained by measuring the fraction of stars accreting disc gas as a function of cluster age.
Several accretion indicators have been discussed in the literature with the optical/UV excess emission from shocked gas at the stellar surface being the most reliable one but also the most challenging to observe (see \cite{2012A&A...548A..56R} for a comparison of accretion indicators). 
On the contrary the H$\alpha$ line at 656.28\,nm is bright, easy to observe with current facilities, and its luminosity correlates well with the optical/UV excess emission 
\cite{2014A&A...561A...2A}, hence its EW and/or width are often used to identify the population of accreting stars. This diagnostic has been used to show that {\bf the fraction of accreting stars declines with cluster age} approaching zero at $\sim$10\,Myr and, assuming an exponential decay, {\bf the e-folding time is 2.3\,Myr}, the same as the dust e-folding time measured at 3-12\,\micron{} \cite{2010A&A...510A..72F} (see previous subsection and Figure~\ref{fig:diskfreq} left panel). The rate at which disc gas accretes onto the star (hereafter, $\dot{M}_{\rm acc}$) is also found to decline with time 
\cite{2010ApJ...710..597S,2010pdac.book..263P}, as expected by viscous evolution, but the spread of  $\dot{M}_{\rm acc}$ at any age is large and the ages of young  stars are poorly known to critically test theoretical predictions.
Overcoming the age issue, a correlation between  $\dot{M}_{\rm acc}$ and $M_{\rm dust}$ has been reported in Lupus \cite{2016A&A...591L...3M}, with a ratio that is roughly consistent with the expected viscous timescale when assuming an interstellar medium gas-to-dust ratio. However, the observed correlation appears to be mainly driven by the underlaying correlation with $M_*$ (Mulders et al. in prep). Indeed, $\dot{M}_{\rm acc}$ is found to scale rather steeply with stellar mass  with a relation close to a power law of two in several star-forming regions: $\dot{M}_{\rm acc} \propto (M_*)^2$ 
\cite{2008ApJ...681..594H,2012A&A...548A..56R,2014A&A...561A...2A,2016A&A...585A.136M}. The steepness of the relation has challenged viscous disc models from early on 
\cite{2006ApJ...648..484H} and has prompted a number of theoretical attempts to explain it. These include Bondi-Hoyle accretion \cite{2005ApJ...622L..61P} (see also \cite{2008AJ....135.2380T}) and dependance on the initial conditions of the parent cloud from which the protoplanetary disc formed \cite{2006ApJ...645L..69D, 2006ApJ...639L..83A}. Recent work has shown that the relation may also  be a consequence of viscous photoevaporative disc clearing driven by X-ray radiation from the central star \cite{2014MNRAS.439..256E}. In this case it was shown that the relation between accretion and stellar mass simply reflects the relation between the X-ray luminosity and the stellar mass, as measured by several surveys including the Chandra Orion Ultradeep Project \cite{2005ApJS..160..401P}. 

Accretion indicators probe the reservoir of disc gas very close to the star, hence it is natural to ask whether gas at tens or hundreds of AU from the star has a similar lifetime. Unfortunately, surveys at infrared and millimeter wavelengths probing those radii are still limited in sample size, but their constraints of a gas lifetime of $\sim$10-20\,Myr are consistent with the value obtained from accretion indicators \cite{1995Natur.373..494Z,2006ApJ...651.1177P,2013PASP..125..477D} (see also Figure~\ref{fig:diskfreq} left panel). In addition, modeling of multiple gas diagnostics from a sample of 15 solar-type stars that lack signatures of accretion but possess optically thin dust emission shows that the gas surface density at 1\,AU has dropped to less than 0.01\% of the minimum mass solar nebula \cite{2006ApJ...651.1177P} in most systems, implying too little gas to even circularize the orbits of terrestrial planets.

The gas clearing time is yet unconstrained observationally as resolved images of gas cavities are just beginning to be possible. However, there may be an indication for gas depletion in spectrally resolved gas lines. After taking into account stellar mass and disc inclination effects, the SED-identified {\it transition} discs have narrower UV H$_2$ flourescent emission lines 
than full discs \cite{2015ApJ...812...41H} and a much weaker or absent broad component\footnote{Optical [OI] broad components have FWHM$>$ 40\,km/s while M-band CO broad components have a FWHM typically larger than $\sim$50\,km/s.} in the optical [OI] forbidden lines \cite{2016ApJ...831..169S} and in the M-band CO rovibrational lines 
\cite{2009ApJ...699..330S,2015ApJ...809..167B}. As transitional objects have a dearth of dust grains in their inner disc and might be, on average, lower accretors (e.g. \cite{2015MNRAS.450.3559N}, but see also \cite{2014ApJ...787..153K} for a possible stellar-mass bias), it should be demonstrated that a reduced gas scale height due to less heating from dust grains and accretion does not result in undetectable broad, i.e. high velocity, emission.
A total of 6 {\it transition} discs have been recently imaged at millimeter wavelengths with ALMA in CO and in the continuum at a spatial resolution sufficient to test if gas is depleted within the known dust cavities. In all cases gas cavities, i.e. where the gas surface density decreases by a factor larger than 10, are found to be smaller than the dust cavity \cite{2016A&A...585A..58V}. 
This finding is taken as evidence for embedded planets because planets can clear their orbits in the gas and trap millimeter grains at the outer edge \cite{2015A&A...573A...9P}. However, other physical processes can also cause a pressure gradient in the gas that will trap dust grains at larger radii compared to the gas cavity. 
It is also worth noting that the main CO isotopologue may not be a good gas density tracer in the inner disc as it readily becomes optically thick, e.g. the case of LkCa15 for which the $^{12}$CO(6-5) transition indicates gas depletion inside the dust disk while the HCO$^+$(4-3) line does not \cite{2016ApJ...833..260D}.
Finally, all discs imaged so far have large dust cavities ($>$25\,AU and up to 140\,AU) and are not representative of the entire population of {\it transition} discs \cite{2016A&A...592A.126V} (see also  Section~\ref{sect:thevsobs}). ALMA observations of the gas content of more typical {\it transition} discs would be extremely valuable. 

\begin{figure}[b]
\centering
\includegraphics[width=1 \textwidth]{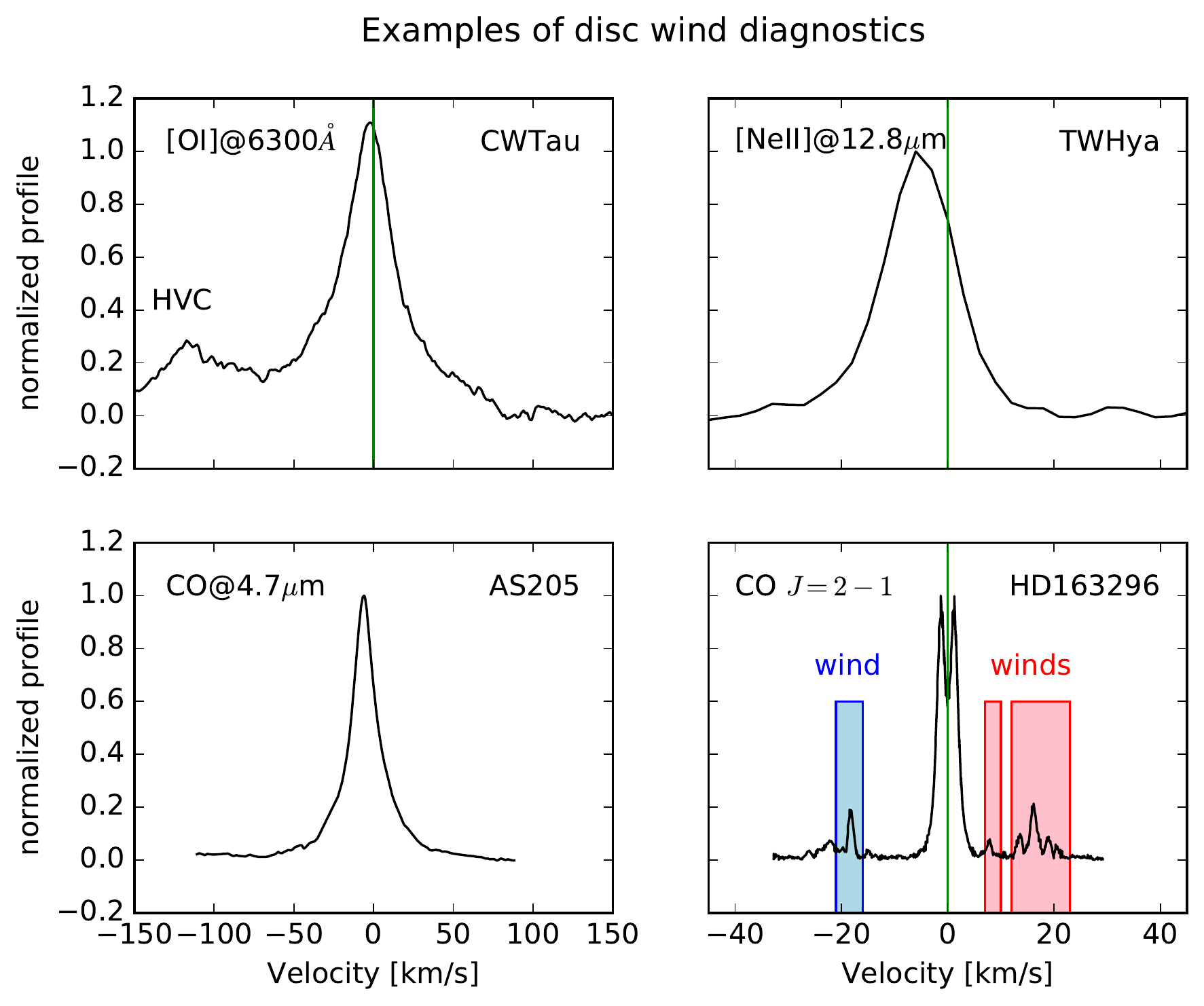}
\caption{Examples of wind diagnostics. The [OI] 6300\,\AA{} profile of CW~Tau is from \cite{2016ApJ...831..169S}, note that the high-velocity component (HVC) is associated with fast jets. The [OI] 5577\,\AA{} transition is weaker than the 6300\,\AA{} but shows a very similar low-velocity profile \cite{2016ApJ...831..169S}. The CO-M band profile of AS205 is from stacked CO rovibrational lines around 4.7\,\micron{} \cite{2015ApJ...809..167B}. The [NeII] 12.8\,\micron{} profile is the mean TW~Hya profile \cite{2011ApJ...736...13P} and the CO J=2-1 profile of HD163296 is from \cite{2013A&A...555A..73K}. All profiles, except that of AS205, are in the stellocentric reference frame and emission shifted in velocity with respect to the stellar velocity is indicative of unbound gas. M-band CO profiles do not have an absolute velocity calibration, indication of a slow wind in AS205 comes from spectro-astrometry \cite{2011ApJ...733...84P}.}\label{fig:exwinds}
\end{figure}

\subsection{Disc winds}
Direct evidence of gravitationally unbound/outflowing material from the disc+star system is achieved by measuring velocity shifts in the emission lines of ionic, atomic, and molecular lines.

Class~I protostars are known to power parsec-long collimated jets that are bright in optical forbidden lines 
\cite{2001ARA&A..39..403R}. Similar gas tracers applied to Class~II CTTs reveal a high-velocity component, typically blueshifted by 30 to 150\,km/s, which forms in micro-jets, a scaled-down version of the Class~I jets 
\cite{2007prpl.conf..231R}. However, typical ratios of jet mass flux to accretion rate are low $\le$0.1 
\cite{2007IAUS..243..203C,2004ApJ...616..998W}, hence jets/outflows do not seem to play a major role in dispersing Class~II discs. More promising are slow ($<$30\,km/s) disc winds, also traced by optical forbidden lines, most frequently by the [OI] line at 6300\,\AA{} 
\cite{1995ApJ...452..736H,2013ApJ...772...60R,2014A&A...569A...5N} (see also Figure~\ref{fig:exwinds} upper left panel). 
It has been recently found that such low velocity emission is present in all TTs with dust discs, even those classified as WTTs, and can be described by a combination of a broad and a narrow component \cite{2016ApJ...831..169S}. Most of the broad component emission arises within $\sim$0.5\,AU and shows the expected disc wind signature, i.e.  larger blueshifted centroid velocities associated with narrower lines and lower disc inclinations (see Figure~27 in \cite{2016ApJ...831..169S} and predicted profiles in Section~\ref{sect:theory}). Such winds must be magnetically driven given that the emitting region is well inside the gravitational radius even for gas at 10,000\,K (see also Section~\ref{sect:theory}). The narrow component is overall less common but always present in transition discs. It traces gas further away ($\sim$0.5-5\,AU) and may be associated with photoevaporative winds as indicated by the blueshift in half of the sample, although the expected trend between blueshifts and disk inclinations is not seen.

Stronger evidence for thermally driven winds comes from mid-infrared high-resolution spectra of [NeII] at 12.8\,\micron. The low-velocity emission has blueshifts between $\sim 2$ and 12\,km/s and FWHM up to $\sim$50\,km/s that are consistent with photoevaporative winds (see Sect.~3.2 in \cite{2014prpl.conf..475A} and references therein). Among these objects it is worth pointing out TW~Hya, a nearby low-mass star with a well known transition disc 
\cite{2016ApJ...820L..40A}. 
Direct evidence of on-going photoevaporation is provided by multi-epoch blueshifted ($\sim$ 5\,km/s) and asymmetric [NeII] 12.8\,\micron{} emission, most of which arises beyond the dust cavity \cite{2011ApJ...736...13P} (see also Figure~\ref{fig:exwinds} upper right panel). All other observables are consistent with the gap being opened by star-driven photoevaporation \cite{2016arXiv160903903E}.

Molecular tracers of outflow activity are common in the protostellar phase. They also probe a low velocity ($<$10\,km/s) component whose intensity decreases while the opening angle increases going from the Class~0 through to the Class~II phase 
\cite{2014prpl.conf..451F}.
This slow wide-angle component, around a much faster jet, is naturally produced in MHD disc winds launched out to several AU 
\cite{2007prpl.conf..277P,2012A&A...538A...2P}. Corroborating this picture, ALMA 
recently mapped a slow CO disk wind around the fast jet from the $\sim$4\,Myr-old Herbig~Ae star HD~163296 \cite{2013A&A...555A..73K} (see also Figure~\ref{fig:exwinds} lower right panel).
 Interestingly, the estimated mass loss rate from the molecular wind is similar to the mass accretion rate onto HD~163296 suggesting that winds contribute significantly to disc dispersal. Like optical forbidden lines, M-band CO ro-vibrational lines also show a broad and a narrow component 
 \cite{2015ApJ...809..167B} with FWHMs slightly  larger than the [OI] 6300\,\AA{} lines suggesting that CO traces gas at smaller disk radii around TTs \cite{2016ApJ...831..169S}. While some profiles are double-peaked as expected if emission arises in a Keplerian discs, others have a single peak with a broad-based emission 
 \cite{2011A&A...527A.119B} reminiscent of the optical forbidden lines. It is yet unclear if the single-peaked M-band CO profiles trace a disc wind as  spectra published so far lack an absolute radial velocity calibration. The spectroastrometric signal detected for two Class II sources (RU~Lup and AS~205N) is intriguing and can be explained by a wide-angle disc wind 
\cite{2011ApJ...733...84P} (see also  Figure~\ref{fig:exwinds} lower left panel for the CO M-band profile from AS205N). In the case of AS~205N, followup high-resolution ALMA observations also show deviations from Keplerian rotation in millimeter CO rotational lines which could be due to a low velocity disk wind or tidal stripping by its companion AS~205S \cite{2014ApJ...792...68S}.



\section{Theories of disc dispersal}\label{sect:theory}

Of all mechanisms that have been brought forward to explain transition-like features in discs, only two can be considered true global disc dispersal mechanisms: accretion and disc winds\footnote{Stellar winds appear to be less efficient than accretion and disc winds 
\cite{2009ApJ...700...10M}}. These processes together shape the evolution of the surface density of the disc and remove material from it by respectively depositing it onto the central object or launching it away from the system. There is strong observational evidence that both processes occur in discs, which has been discussed in the previous sections. In fact accretion and mass loss via a wind compete in determining the surface density of the disc at a given radius. To zeroth order a gap will be formed in a disc if the wind is faster at removing material from a given region than accretion is at pushing new material into it. 

It is classically assumed that for the largest part of their lives, the evolution of the surface density of discs can be well described by simple viscous theory 
\cite{1998ApJ...495..385H,1974MNRAS.168..603L}. These models predict a slow, homogeneous dispersal of the disc. Observations, however, show that the dispersal is not a continuous process: after having evolved slowly 
for a few million years, discs regularly seem to disappear quite abruptly, implying a disc dispersal timescale about 10 times faster than the global disc lifetime (Section~\ref{sect:obs}a). These observations have motivated the development of {\bf theoretical models able to match this two-timescale and inside-out dispersal modus}. Photoevaporation from the central star is currently accepted as an important  player in the late evolution of discs and has seen several dedicated theoretical efforts 
\cite{2001MNRAS.328..485C, 2006MNRAS.369..216A, 2006MNRAS.369..229A, 2008ApJ...688..398E, 2010MNRAS.401.1415O, 2011MNRAS.412...13O, 2012MNRAS.422.1880O, 2009ApJ...705.1237G}. All models of photoevaporation show that radiation from the central star heats the disc atmosphere, where a thermal wind is established. The mass loss rate of the wind must exceed the accretion rate in the disc for dispersal to set in. Young discs accrete at a vigorous rate, which naturally decreases as time goes by, until, after a few million years accretion rates fall to values smaller than the wind rates, allowing photoevaporation to take over the further evolution of the disc. Once the dispersal sets in the disc is then quickly eroded from the inside out (Section~\ref{sect:theory}b). This scenario is based on the assumption that the mass accretion rate is radially constant throughout.

While the community seems to agree on this broad brush picture, quantitatively speaking, the dispersal mechanism is still largely unconstrained. In fact, photoevaporation by energetic radiation from the central star is not the only way to produce a disc wind. Extended magnetically-launched disc winds appear to be necessary to explain the properties of jets in young stars 
\cite{2006A&A...453..785F} and models have been developed since the 1980s 
\cite{1982MNRAS.199..883B}. However, only recently non-ideal MHD effects have been incorporated, albeit with computational costs that have mostly limited the calculation to small regions of the disc. Some of these recent models, based on local simulations, suggest that even a weak vertical magnetic field can launch a wind 
\cite{2013ApJ...769...76B}. The wind can be so vigorous as to compete with photoevaporation for the dispersal of the disc and may even provide an efficient channel to remove angular momentum from the material in the disc, hence driving the accretion process. Global MHD discs simulations have until very recently only been possible in the ideal limit \cite{2014ApJ...784..121S,2015ApJ...801...84G}, or including only Ohmic diffusion \cite{2013ApJ...765..114D, 2017ApJ...835..230F}. Generally the Ohmic term is sub-dominant to the Hall term in the inner disk (inside ~10 AU), and sub-dominant to ambipolar diffusion further out. 

The most recent simulations in the ideal limit  \cite{2017arXiv170104627Z}, which improve on resolution and convergence compared to previous work, do not show winds that significantly contribute to the evolution of the surface density of the disc under the assumption made.
At the time of writing only one set of global simulations of protoplanetary discs including all three non-ideal MHD effects (ohmic and ambipolar diffusions, and the Hall drift) has been performed \cite{2016arXiv161200883B}. Discs are found to accrete (in which case a disc wind is also present) only for given configurations of the large-scale magnetic field, which thus remains an important uncertainty. 

The nature of disc accretion and the driving mechanisms behind disc winds are currently a rapidly developing and very active area of research. An attempt at summarising the state-of-the art at the time of writing follows. 

\subsection{Accretion} 

Accretion in discs is often phenomenologically described by the so-called $\alpha$ prescription \cite{1973A&A....24..337S}, where the paramater $\alpha$ is used to describe the viscosity of the material, and it represents the ratio of stress over pressure. The source of the viscosity is however still a matter of debate. The standard paradigm for circumstellar disc accretion invokes the magneto-rotational instability (MRI) to drive turbulence 
\cite{1991ApJ...376..214B}. The conditions to trigger MRI is that the gas is sufficiently ionised, the disc is weakly, but non-negligibly magnetised and that the angular frequency decreases with radius. While the last two conditions are always satisfied in discs, there are uncertainties about the ionisation structure of the gas. 
It is indeed often pointed out in the literature that large disc regions may be actually MRI-inactive, due to poor coupling of the gas to the magnetic fields. In general, regions lying under large column densities of material in the inner disc, may be virtually neutral, the so called ``dead zones'' 
\cite{1996ApJ...457..355G}, where MRI turbulence is practically absent. Accretion in these cases is thought to happen in the outer layers surrounding the dead zones. These regions, however, may be dominated by non-ideal MHD effects (e.g. ambipolar diffusion, Ohmic diffusion) which may make MRI transport very inefficient 
\cite{2014ApJ...783...14T}. 

Recent non-ideal MHD simulations 
\cite{2013ApJ...769...76B,2016arXiv161203912B} show that in the presence of even a weak net vertical magnetic field threading the disc, MRI is completely suppressed. As a result however a vigorous magnetocentrifugal wind is launched in these simulations. The wind in this case also removes angular momentum from the disc and can thus drive accretion. Using the results of local simulations a 1+1D global model of a disc evolving under the influence of MHD winds can be constructed \cite{2013ApJ...769...76B, 2016ApJ...821...80B}. Based on this approach recent work has found that accretion is completely dominated by winds in most regions of the disc, with the MRI perhaps only playing a part in the outer regions, e.g. at R$\gtrsim50~AU$ for the fiducial model  \cite{2016ApJ...821...80B}. In this scenario the accretion and wind mass loss processes are strongly coupled by the field strength which drives both. While the suggestion of MHD winds driving accretion in discs is certainly tantalising, it is currently too early to assess its validity. In particular, all attempts at constructing global models for the evolution of discs rely on a number of important assumptions about how critical parameters scale with radius \cite{2016ApJ...821...80B} and are sensitive to the amount of magnetic flux assumed to thread the disc. 

Regardless of the detailed process(es) that provide the viscosity needed to drive accretion in discs, if the phenomenologically formulated alpha-description is roughly correct, then one expects the surface density to decrease as a power law of time and the radius to increase, as a result of viscous draining and spreading, i.e.  $\Sigma_{disc}(t) \sim \dot M_{disc}(t)~t^{-1.5}$
\cite{1998ApJ...495..385H}. The implication is that viscously evolving discs should become progressively faint simultaneously at all wavelengths as a function of time. Due to the temporal power law, the evolution should become progressively slower, such that discs should spend most of their lifetime in an homogeneously draining transition phase. To date there is no convincing observational evidence of any homogeneously draining disc. Current data support instead a fast dispersal phase which proceeds from the inside out (Section~\ref{sect:obs}a). Viscosity is thus not the end of the story, rather an additional dispersal process must take over at advanced stages of disc evolution. Mass loss via a disc wind, driven by photoevaporation by the central star, likely plays an important role, such that the evolution of the surface density of a disc switches from an accretion-dominated to a wind-dominated regime, marking the beginning of  the final, rapid disc dispersal phase. However, if MHD winds dominate angular momentum transport in a significant part of the disc, the evolution of the surface density and of the mass accretion rate is unlikely to be well described by an alpha-formalism 
\cite{2015ApJ...815..112K, 2016ApJ...821...80B} . While current observations are reported to be broadly consistent with the alpha-disc scenario, they cannot unequivocally prove it, as discussed in more detail in Section~\ref{sect:obs}a.

\subsection{Photevaporative/thermal winds}

\begin{figure}[b]
\centering
\includegraphics[width=0.7 \textwidth]{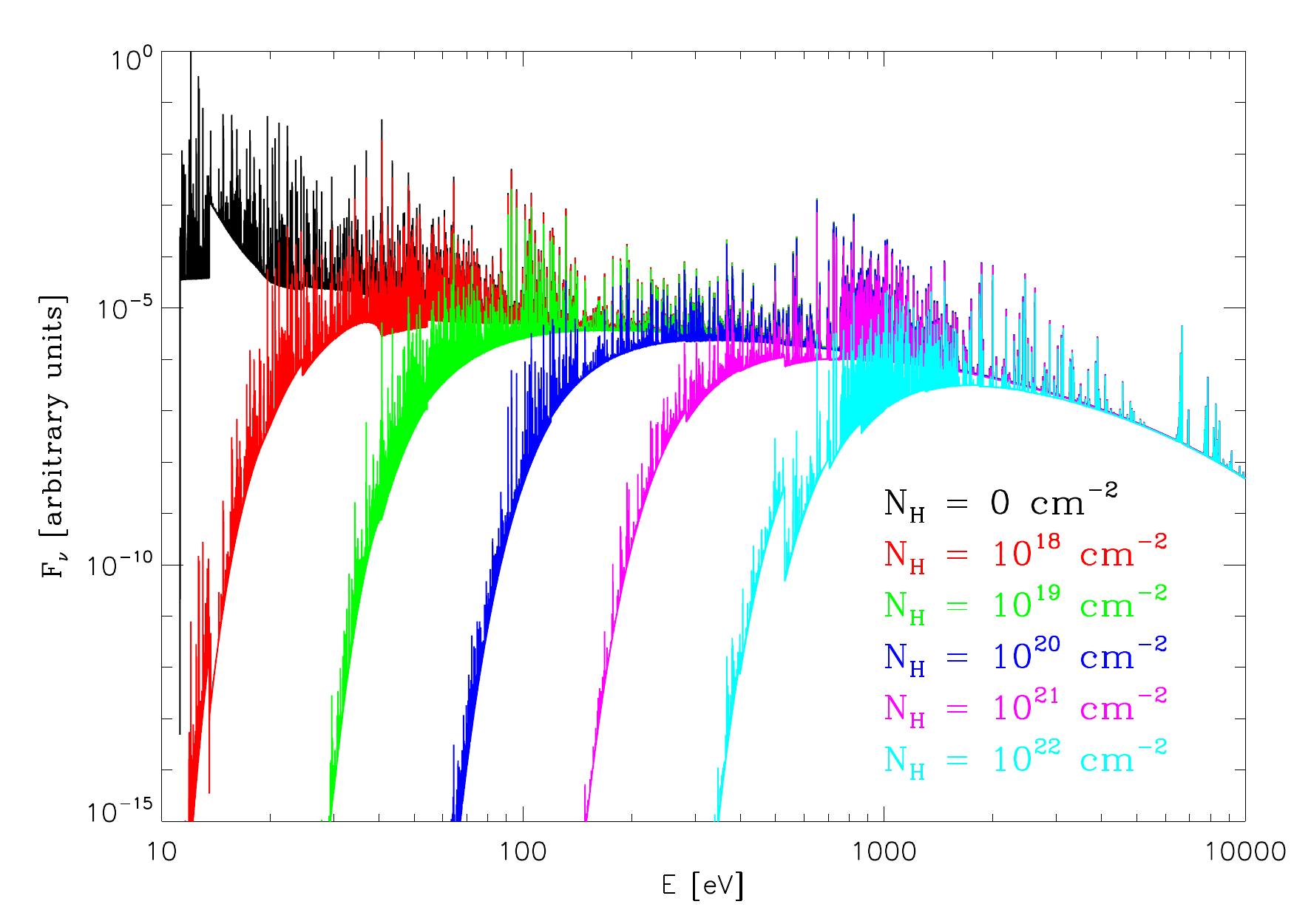}
\caption{Attenuation of a model coronal spectrum through columns of neutral hydrogen. The spectra are in arbitrary units. The N$_H$ = 0 curve refers to the unattenuated model. Figure adapted from \cite{2009ApJ...699.1639E}. }\label{fig:attenuation}
\end{figure}

Theoretical work on photoevaporation has been recently reviewed by a number of authors 
\cite{2011ARA&A..49..195A, 2014prpl.conf..475A, 2016SSRv..205..125G}. We will only summarise the main mechanisms here, focusing on the current open questions and uncertainties, as well as on the specific predictions and the predictive power of current models. 

Radiation from the central star penetrates the disc atmosphere and deposits energy in (i.e. heats) the gas. To zeroth order a thermal wind is established if the temperature of the gas at a given location becomes higher than the local escape temperature. Far-Ultraviolet (FUV, $6eV \lesssim E \lesssim $13.6eV ), Extreme-Ultraviolet (EUV, $13.6 eV \lesssim E \lesssim 100 eV$  ), and soft X-rays ($0.1keV  \lesssim E \lesssim 2 keV$) are all able to efficiently heat the gas in the disc atmosphere. As shown in Figure~\ref{fig:attenuation}, the penetration depth of the photons depends however strongly on wavelength, with the EUV being absorbed out completely within a column of neutral hydrogen of $\sim10^{20}$\,cm$^{-2}$ 
\cite{2009ApJ...703.1203H}. FUVs and soft X-rays have larger penetration depths, of order 10$^{21}$, 10$^{22}\,cm^{-2}$, depending on the abundance of small dust grains and PAHs in the case of the FUV and the spectral shape in the case of X-rays \cite{2009ApJ...699.1639E}. High energy X-rays ($E\gtrsim 2keV$) can penetrate very large columns, but they do not heat the dense gas efficiently enough to drive a wind.
Again to zeroth order, the surface mass loss rate at a given disc radius is given by $\rho_b*c_s$, where $c_s$ and $\rho_b$ are the sound speed and the density of the gas at the base of the flow, respectively. Given that the temperature of the gas only enters as a square root in the calculation of the sound speed, and that $\rho$ varies steeply as a function of vertical depth $z$ in the disc, the surface mass loss at a given radius is much more sensitive to the location of the base of the flow than to the temperature of the gas 
\cite{2010MNRAS.402.2735E}. The FUV and soft X-rays are thus expected to drive the strongest winds 
\cite{2008ApJ...688..398E, 2009ApJ...699.1639E, 2010MNRAS.401.1415O, 2011MNRAS.412...13O, 2012MNRAS.422.1880O, 2009ApJ...705.1237G}. There is still a debate in the literature as to what type of radiation may be the main driver of the wind, although EUV photons alone can be excluded in a few sources \cite{2004A&A...417..793P}, see also Section~\ref{sect:theory}d. This is a fundamental question as the mass-loss-rates implied by the different models can differ by orders of magnitudes. Mass loss rates crucially determine the timescales of disc dispersal for given initial disc conditions.

Hydrodynamical simulations show that a wind driven by EUV radiation only would result in mass-loss rates of order 10$^{-10} M_{\odot}/yr$ for a solar-mass star irradiated by photon luminosity of $\phi \sim 10^{41}$ phot/sec \cite{2006MNRAS.369..229A}. Radiation-hydrodynamic models including both X-ray and EUV radiation (XEUV),  yield instead mass-loss rates of order 10$^{-8} M_{\odot}/yr$ for a disc around a solar type star irradiated by a star with X-ray luminosity of $L_X \sim 10^{30} erg/sec$ \cite{2010MNRAS.401.1415O, 2011MNRAS.412...13O, 2012MNRAS.422.1880O}. The scaling of the mass loss rate in the two cases is also very different: $\dot M \propto \sqrt{\phi}$ for pure EUV and $\dot M \propto L_X$ for the XEUV case. Thus the mass-loss-rates predicted by the XEUV model for solar-type stars may vary strongly depending on the X-ray property of the central object. X-ray surveys of young stars show approximately two orders of magnitude scatter in the X-ray luminosities of young solar-mass stars. The contribution of short term variability to the observed scatter is minor \cite{2005ApJS..160..390P}, with intrinsic differences in stellar rotational velocities or internal structure being instead the dominant factor. The same scatter will then be reflected in the mass-loss-rates, and thus in the expected lifetimes of their discs. The same is not true for an EUV-only scenario where the dependence of the mass-loss-rate on the EUV luminosity is much weaker. Thermochemical models including EUV, FUV and X-ray heating, also find mass-loss-rates of order 10$^{-8} M_{\odot}/yr$ for solar-type stars  \cite{2009ApJ...705.1237G}. 
These models, however, do not perform a hydrodynamical calculation to obtain a solution for the wind. The mass loss rate from the 1+1D hydrostatic equilibrium models at a given radius is estimated using the approximation discussed above ($\rho_b*c_s$), where the location of the base of the flow is chosen so that $\rho_b*c_s$ is maximised. It is difficult to estimate the uncertainty introduced by the method, thus a comparison with hydrodynamical models  \cite{2006MNRAS.369..229A, 2010MNRAS.401.1415O, 2011MNRAS.412...13O, 2012MNRAS.422.1880O} is of limited relevance.

The wind profile, which determines the region of the disc that is most affected by photoevaporation, is also very different in each scenario 
\cite{2011ARA&A..49..195A, 2014prpl.conf..475A}. The X-ray profile is more extended than the EUV profile, which predicts mass loss only from a vary narrow range of disc radii, centred at the gravitational radius. The FUV model is again very different, showing mass loss from the outer regions of the disc and predicting in some cases an outside-in mode of dispersal. 
The detailed profile of the photoevaporative wind has important consequences on the formation and migration of planets in the wind. As an example, it has recently been shown that changing the wind profile yields completely different distributions for the semi-major axes of giant planet in otherwise equal populations of discs losing mass globally at the same rate \cite{2015MNRAS.450.3008E}. 

\begin{table}
\noindent
\begin{tabular}{lccc}
\hline
Ingredients           & Alexander et al. (2006) & Owen et al. (2010) & Gorti et al. (2009) \\ 
\hline
Hydrodynamics         &    Yes                  &     Yes            &  No                 \\
Thermal Calculation   &    No                   &     Yes            &  Yes                \\
Chemistry Calculation &    No                   &     No             &  Yes                \\
FUV heating           &    No                   &     No             &  Yes                \\
EUV heating           &    Yes                  &     Yes            &  Yes                \\
X-ray heating         &    No                   &     Yes            &  Yes                \\
\hline 
\end{tabular}
\caption{Summary of main ingredients included in current photoevaporation models.}
\label{tab:models}
\end{table}

While the physics of photoevaporation is reasonably well-understood, all current models are somewhat incomplete. The EUV-model focuses on hydrodynamics and assumes isothermal gas 
\cite{2004ApJ...607..890F, 2006MNRAS.369..229A,2006MNRAS.369..216A}, hence being limited to the EUV-ionised layer only, which yields roughly constant temperatures around a value of 10,000\,K. The available FUV models focus on chemistry, but do not perform hydrodynamical calculations \cite{2009ApJ...705.1237G, 2015ApJ...804...29G}.
Current radiation hydrodynamic calculations of X-ray driven winds use realistic gas temperatures obtained from X-ray photoionisation calculations \cite{2009ApJ...699.1639E}, but they do not include chemistry and ignore the dust phase \cite{2010MNRAS.401.1415O, 2011MNRAS.412...13O, 2012MNRAS.422.1880O}. Indeed none of the existing models take into account dust evolution in the underlying disc and entrainment of grains in the wind self-consistently. Table ~\ref{tab:models} summarises the main ingredients included by the models, as discussed above. There are further important differences in the implementation of the various heating channels, as well as on assumptions made by the different codes. A full technical discussion of the different choices adopted and their influence of the derived mass loss rates and profiles is still at this stage, since not all ingredients are included in all models. A benchmarking exercise would be nevertheless useful at this stage in order to converge on a roadmap for future development.

\subsection{MHD winds}

Before we begin our discussion on magnetically supported/driven winds, it is worth noting that there exist a large body of literature on MHD models developed to explain outflows from young stars, particularly those of Class I. One set of these models were based on disc/magnetosphere interactions and mostly deal with the ideal MHD region of the disc 
\cite{1994ApJ...429..781S,2006A&A...453..785F}. Including a comprehensive discussion of these early works is beyond the scope of this review. Here we will focus on MHD winds only in the context of disc dispersal. 

As mentioned in Section~\ref{sect:theory}a, MRI \cite{1991ApJ...376..214B} is currently considered a likely mechanism by which angular momentum in discs is redistributed to allow accretion of matter through the disc and onto the central object. Classically, magneto hydrodynamical (MHD) simulations in a local shearing box have been used to study the properties of turbulence driven by the MRI 
\cite{1995ApJ...440..742H}. More recently,   vertically stratified local shearing box simulations have been used to investigate the possibility of MRI-driven disc winds \cite{2009ApJ...691L..49S, 2010ApJ...718.1289S}. Their model differed from previous work, that used initially toroidal and zero-net vertical flux magnetic fields 
\cite{2000ApJ...534..398M}, by assuming vertical magnetic fields and outgoing boundary conditions. This was motivated by the assumption that discs may be threaded by net vertical magnetic fields that are connected to their parental molecular cloud. These authors find that vigorous winds can be driven by the MRI, which may disperse the discs with timescales as short as 4,000\,yr at 1\,AU and 6 $\times$ 10$^5$ yr at 30\,AU. These rather short timescales, which are clearly in contrast with disc observations, do not account for global viscous accretion, which would slow down the process. Very recent work expands on these results to present 1D models of disc evolution including the effects of viscous heating, in addition to the loss of mass and angular moment by the disc wind \cite{2016ApJ...821....3M}. The focus of this work is more on the early stages of evolution, when accretion heating is important and can give rise to density structures which show a positive radial slope for the surface density in the inner disc regions. While this may have important implications for planetesimal formation and migration models, its relevance to the dispersal of discs is limited. 
The latest global ideal MHD simulation performed, at the time of writing this review, is for thin accretion disks threaded by net vertical magnetic fields  \cite{2017arXiv170104627Z}. This work suggests that only very weak and episodic disc winds can be driven and these are not efficient at carrying angular momentum. 

As mentioned in the previous section, however local non-ideal MHD simulations suggest  instead that in the presence of a weak net vertical magnetic field, MRI is completely suppressed, while a strong magnetocentrifugal wind is launched, which carries away disc angular momentum so efficiently to account for the measured accretion rates \cite{2013ApJ...769...76B}. 
1D, vertically integrated, disc evolution models including angular momentum redistribution and wind angular momentum loss, have been constructed using a parameterisation of the wind stress parameter from the vertically stratified ambipolar disc simulations \cite{2013ApJ...775...73S, 2013ApJ...778L..14A}. The simulations modelled MRI turbulence at 30 AU in an FUV-ionised disc and lead to the conclusion that, depending on the value of the initial net flux, discs may in some cases undergo a two-timescale dispersal behaviour, similar to that achieved by photoevaporation  \cite{2013ApJ...778L..14A}. For low initial magnetic field strengths discs are almost inviscid and very long-lived. However, even in the cases where the initial field is sufficient to instigate rapid disc dispersal, it is not possible to predict what kind of transition disc morphology would be obtained or whether the disc would then disperse from the inside-out, as observations suggest (see Section~\ref{sect:obs}a). The behaviour of MHD dispersal models depends upon the radial evolution of the net field and how this couples to disc evolution, which is currently unknown \cite{2013ApJ...778L..14A}.
A simplified (1+1D) approach, motivated by numerical simulations \cite{2013ApJ...769...76B} has been used to model MHD wind driven disc evolution \cite{2016arXiv161203912B, 2016ApJ...821...80B}. This study shows that, under the assumptions made, wind-driven accretion and mass loss, rather than MRI, dominate disc evolution. Later work also tries to incorporate some thermodynamical effects using simplified prescriptions for disc temperature and the depth of the FUV penetration \cite {2016ApJ...821...80B}. In this work it is found that, while FUV penetration can have significant effects on the wind-driven accretion rates and fractional wind mass loss rates, the key parameter controlling the disc evolution timescales in their models is the amount of magnetic flux threading the disc. This depends both on the initial strength on the magnetic field and its evolution with respect to the surface density evolution of the disc. This is in agreement with previous assessment \cite{2013ApJ...778L..14A}, and it shows that understanding the behaviour of magnetic fields in discs is key to predicting the impact of MHD winds in the evolution and dispersal of discs. While direct observations of magnetic fields in planet-forming discs remain challenging 
\cite{2014Natur.514..597S, 2015ApJ...809...78K}), disc winds are commonly observed (Section~\ref{sect:obs}b) and their properties may be used to indirectly constrain magnetic fields.

\subsection{Prediction of wind structures}

Gas flowing into a wind has a very clear non-Keplerian kinematical signature, which can be observed in the profile of emission lines that are produced within it. 

\begin{figure}[h]
\centering
\includegraphics[width=1 \textwidth]{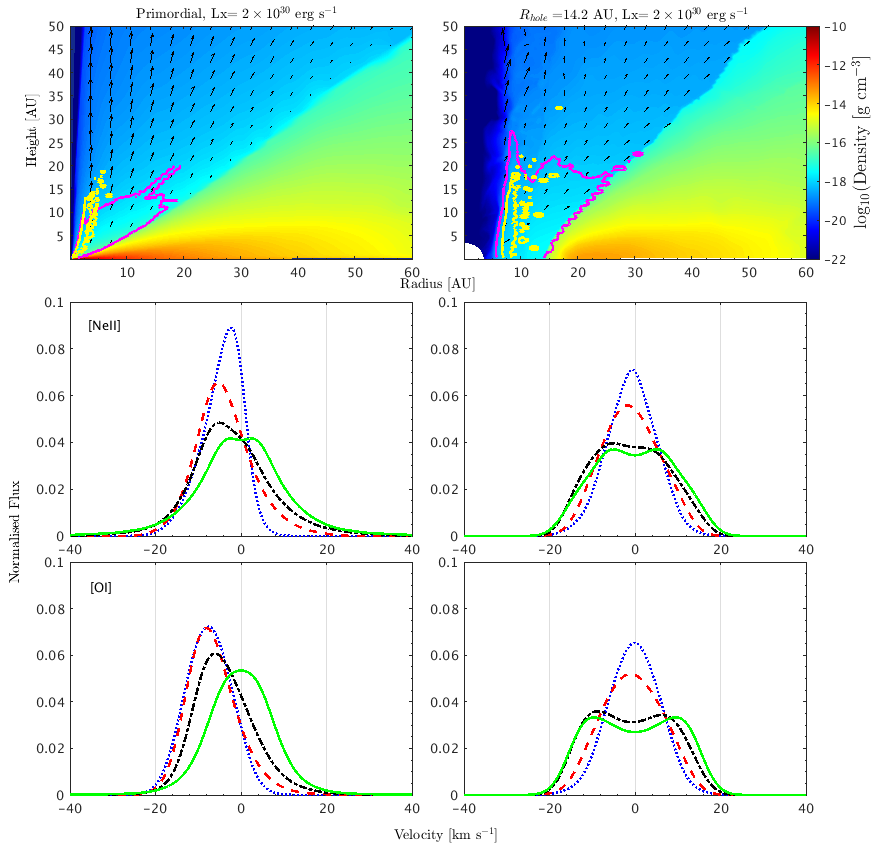}
\caption{Re-rendering of the radiation-hydrodynamics calculations of a full photoevaporating disc (left panels) and a disc with a cavity of 14 AU (right panels)  \cite{2010MNRAS.406.1553E}. The top panels show density maps, where superimposed is the location of the 85\% emission region of the OI 6300A line (yellow contour) and NeII 12.8$\mu$m line (purple contour), the arrows show the direction and magnitude of the velocity vector.
The lower panels show the emission line profiles for  the same lines at four different disc inclinations: 0$^o$ (blue dotted line), 30$^o$ (red dashed line),  60$^o$ (black  dash-dot line) and 90$^o$ (green solid line line)}\label{fig:profiles}
\end{figure}

The physics of photoevaporative winds is relatively well understood and detailed calculations of wind structures, that will improve the ability of models to confront observations, are slowly maturing. Examples of predicted emission line profiles for four different orientations of a full disc (left panels) and one with an inner cavity of approximately 14AU (right panels) are shown in Figure~\ref{fig:profiles}, which is a re-rendering of previous radiation-hydrodynamics calculations \cite{2010MNRAS.406.1553E}. The top panels show density maps, where superimposed is the location of the 85\% emission region of the [OI] 6300A line (yellow contour) and [NeII] 12.8$\mu$m line (purple contour), the arrows show the direction and magnitude of the velocity vector.
The lower panels show the emission line profiles for  the same lines at four different disc inclinations: 0$^o$ (blue dotted line), 30$^o$ (red dashed line),  60$^o$ (black  dash-dot line) and 90$^o$ (green solid line line). Note how the wind emission dominates for close to face-on orientations (0$^o$), leading to a large blueshift and a small FWHM, as seen in some disc wind diagnostics (Section~\ref{sect:obs}b). 

On the other hand, MHD wind models that try to predict the mass and angular momentum loss profiles and flow topology from the disc rely on a number of assumptions about how critical parameters scale with radius 
\cite{2016arXiv161203912B}. Furthermore, as discussed in the previous section, these models are extremely sensitive to details of the magnetic field strength, topology and evolution that are poorly known. As such, a detailed comparison with candidate wind diagnostics is still lacking. However, one important difference that can already be identified using current models of photoevaporative and MHD winds is that the latter are likely to include an unbound higher velocity component launched close to the star ($\lesssim 1-2$\,AU). Photoevaporative outflows, instead, drop off rapidly in strength inside a radius of $\sim 0.2 G M_* / c_S \sim 2 AU$, where $c_s$ is the sound speed \cite{2013ApJ...778L..14A}. 
Therefore, while the identification of an unbound component launched close to the star immediately points to MHD winds \cite{2016ApJ...831..169S} (see also  Section~\ref{sect:obs}b), slower winds driving the mass loss in the 2-10\,AU regime may be MHD or thermal in nature.

As observations of disc winds have grown in the past years (Section~\ref{sect:obs}b), they also revealed that multiple diagnostics are needed to determine wind mass loss rates. For instance,  predicted [NeII] 12.8\,\micron{} line profiles for an EUV-driven wind \cite{2008MNRAS.391L..64A} and an X-ray-driven wind \cite{2010MNRAS.406.1553E} are very similar although the mass loss rates differ by two orders of magnitude. This is because the location of the base of the flow is much deeper in the disc in the X-ray case (see Section~\ref{sect:theory}b). 
However, combining this diagnostic with an upper limit on the EUV luminosity reaching the disk from the cm emission in excess of the thermal dust emission \cite{2012ApJ...751L..42P,2013MNRAS.434.3378O} rules out EUV photoevaporation in at least three cases pointing to mass loss rates larger than $10^{-10}\,M_{\odot}/yr$ \cite{2014ApJ...795....1P}. 
Similarly, in the context of thermal wind models, the [OI] 6300\,\AA{} line cannot be used alone to infer the underlying mass loss rate \cite{2016MNRAS.460.3472E}. This is because, if it is collisional excited, it has a strong temperature dependence imposed by the Boltzmann term in the emissivity and mostly traces the hot layer of the wind, not the bulk. Molecular lines should be more sensitive to the mass loss rate since they sample a significant area of the wind launching regions. However, the exploitation of molecular tracers is currently severely hampered by the lack of a suitable hydrodynamic wind model coupled to chemistry and to dust evolution models (which dominate the opacity in the wind) to interpret the observations. While a number of chemical models exist of the deeper, denser regions of discs, no model is currently available for the optically thinner disc winds. 
Current detailed chemical calculations which extend to the disc atmosphere use a hydrostatic disc model analysed in a 1+1D fashion \cite{2009ApJ...705.1237G}.  
Some studies of chemistry along MHD disc wind streamlines for Class 0-II have been presented however and show promising results \cite{2012A&A...538A...2P, 2016A&A...585A..74Y}. In the future one can thus hope to use similar techniques coupled to a more generalised approach for the calculation of disc wind to better exploit molecular observations to  constrain the wind driving mechanism and directly measure important wind properties.  

\section{Observations vs Theory: The true population of dispersing discs}\label{sect:thevsobs}
\begin{figure}[h]
\centering
\includegraphics[width=1 \textwidth]{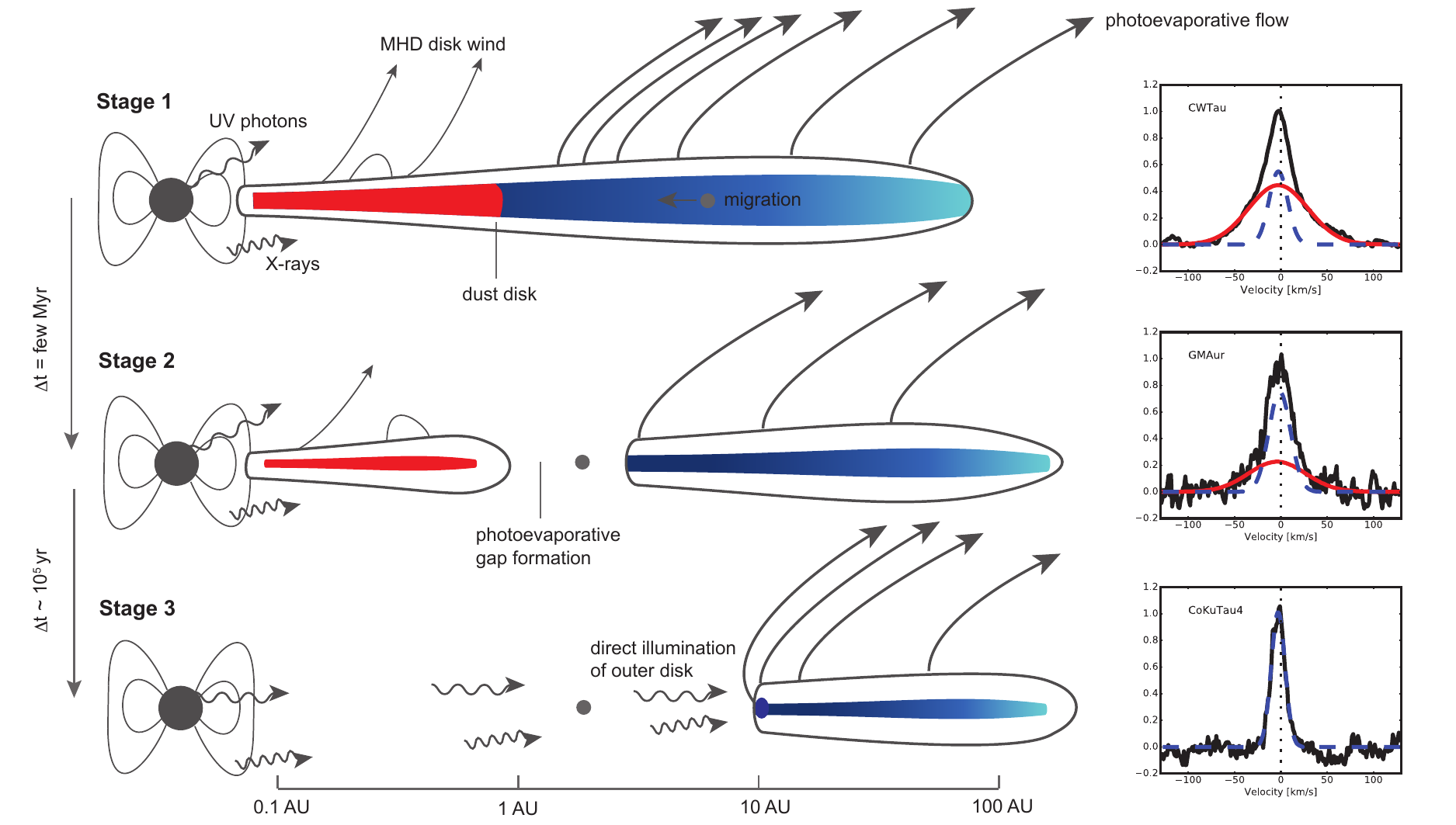}
\caption{The three main stages of disk evolution and dispersal (adapted from a previous review \cite{2014prpl.conf..475A}). The right panels show sample low-velocity [OI] 6300\,\AA{} emission \cite{2016ApJ...831..169S} tracing a possible evolution in disk winds.}\label{fig:ddisp}
\end{figure}

Before comparing observations and theory to identify discs that are truly caught in the act of dispersing, we briefly summarize the salient results discussed in previous sections. We also schematically illustrate the three main stages of disc evolution and dispersal in Figure~\ref{fig:ddisp}.
Disc dispersal processes, accretion and winds, act throughout the disc lifetime as evinced from observations of objects in different SED classes.
Once accretion becomes undetectable most of the dust and gas at large disc radii are also dispersed.
Disc dispersal in low-mass star-forming regions appears to occur from the inside-out as indicated by the detections of dust cavities and hints of gas cavities and gaps. There is also tentative evidence for the evolution of disc winds, with the possible disappearance of MHD winds as discs evolve (see right panels in Figure~\ref{fig:ddisp}).
The transition from disc-bearing to disc-less is fast, $\sim$10\% of the disc lifetime. 

While theoretical models combining accretion and disc winds are consistent with these basic observations and reproduce the lifetime and dispersal of discs, several authors have pointed out that the theory fails to explain a subset of the observed transition discs 
\cite{2016PASA...33....5O}.
But are all transition discs truly caught in the act of dispersing?

\begin{figure}[b]
\centering
\includegraphics[width=1 \textwidth]{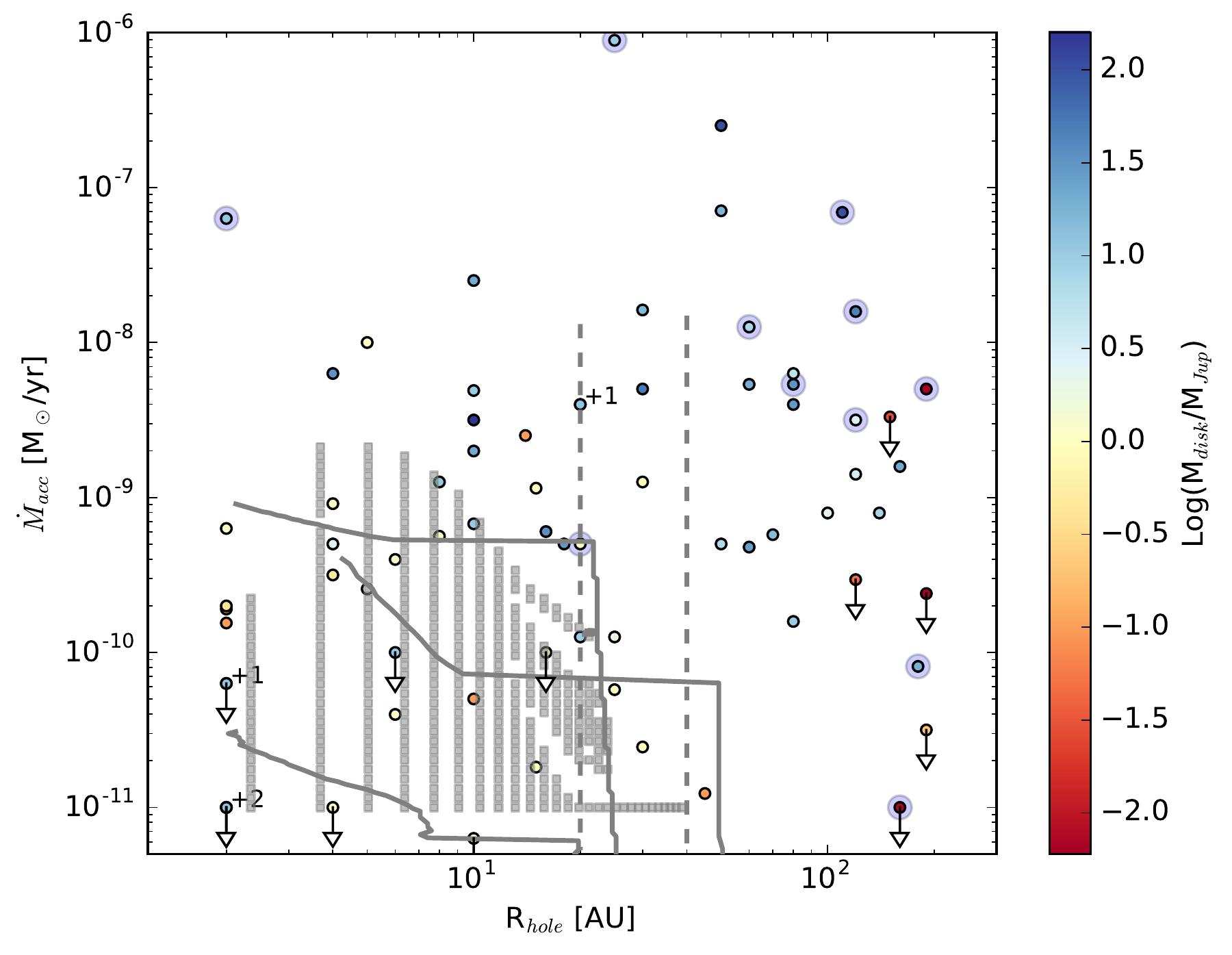}
\caption{Mass accretion rate versus disk hole size. Circles are for observed star-disk systems. The color-coding shows total disk masses estimated from SED fitting, see main text. Sources surrounded by a light blue circle have spectral types G and earlier, note that they are mostly high accretors with large disk holes. Grey squares are snapshots of EUV- plus X-ray-driven photoevaporating discs \cite{2011MNRAS.412...13O} while grey lines are evolutionary tracks for the same photoevaporating discs with an embedded giant planet \cite{2013MNRAS.430.1392R}. Numbers next to a source indicate how many additional sources are at that location, see also footnote number 4.}\label{fig:MaccRhole}
\end{figure}

We make use of the largest classification of transition disc candidates  \cite{2016A&A...585A..58V}, which is based on {\it Spitzer} and complementary long-wavelength photometry, to identify different populations of transition discs in the mass accretion rate ($\dot{M}_{\rm acc}$)
vs hole size ($R_{\rm hole}$) plot 
\cite{2012MNRAS.426L..96O}.
Stellar parameters are also provided in this work \cite{2016A&A...585A..58V}, including spectral type and luminosity (Table A.2), an estimate of $R_{\rm hole}$ from SED modeling, which is insensitive to $R_{\rm hole}\le 1$\,AU, and total disc masses (assuming a gas-to-dust ratio of 100), also obtained from SED modeling (Table A.3). From this sample of $\sim$200 transition disc candidates we identified a subsample of 72 objects with literature mass accretion rates including upper limits and estimated dust holes larger than 1\,AU. This subsample is plotted in Figure~\ref{fig:MaccRhole} and values used for the plot, including references, are provided in Table~A1\footnote{Note that there are sources with the same $\dot{M}_{\rm acc}$ and $R_{\rm hole}$, only 68 individual points can be recognized in Figure~\ref{fig:MaccRhole}. These sources are: i. J163115.7-24340 and DoAr28, Log($\dot{M}_{\rm acc}$)=-8.4 and $R_{\rm hole}$=20\,AU; ii. J162802.6-23550, 03442156+321509, and J130455.7-7739, Log($\dot{M}_{\rm acc}$)$<-11$ and $R_{\rm hole}$=2\,AU; iii. ASR118 and IC348-67, Log($\dot{M}_{\rm acc}$)=-10.2 (non-detection for the second source) and $R_{\rm hole}$=2\,AU.}. Objects are color coded by disc mass, spectral types G and earlier are surrounded by a light blue circle, and upper limits in $\dot{M}_{\rm acc}$ have a down-pointing arrow. We decided to highlight early spectral type stars because they appear to be mostly high accretors surrounded by disks with large dust holes.
Grey squares are snapshots of models of EUV- plus X-ray-driven photoevaporating discs  \cite{2011MNRAS.412...13O} while grey lines are evolutionary tracks for the same photoevaporating discs with an embedded giant planet \cite{2013MNRAS.430.1392R}. The solid and dashed grey lines correspond to two different assumptions for the dust dynamics in the inner disc at the time of gap opening. Solid lines represent models where the dust in the inner disc continues to be replenished from the outer disc after the formation of the gap. In this assumption, small dust grains (in the Epstein regime) are able to move across the gap together with gas. The vertical dashed lines corresponds to models where the planet is assumed to trap all grains in the outer disc, such that as soon as gap forms, the disc is observed to have a dust cavity of size equal to the radius at which the gap was opened by the planet. These latter models also assume that grains in the inner disc drain onto the stars on much shorter timescales than the gas in the inner disc. 
These models illustrate  the range of possible transition discs predicted by photoevaporation. Note how embedded planets can open larger dust gaps but also reduce the transition time 
\cite{2013MNRAS.430.1392R}.


Twenty nine out of 72 discs ($\sim$40\%) fall  in the area covered by grey squares and within the solid grey lines and, as such, are consistent with cavities being opened by photoevaporation. If these are the only discs caught in the act of dispersing, the disc dispersal timescale is about 40\% of that measured from counting all transition discs.
However, this number should be taken as a lower limit as upper limits on mass accretion rates are more rarely reported in the literature than mass accretion rate measurements. Overall these results confirm previous reports that the dispersal timescale is short, of order 10$^5$\,yrs. Once the inner disc has drained onto the central star, Stage 3 in Figure~\ref{fig:ddisp}, theoretical models predict a roughly equal amount of time to consume the outer disc 
\cite{2011MNRAS.412...13O}. At this stage, the star is not accreting (WTT), the hole would be anywhere from a few to $\sim$100\,AU, and, assuming no grain growth, there should be still enough mass in small grains for the disc to be optically thick. Thirteen out of the 72 discs ($\sim$20\%) in Figure~\ref{fig:MaccRhole} could be in this stage, a factor of two lower than the accreting transition discs consistent with photoevaporation. Again, this fraction should be taken as a lower limit because mass accretion rates are often not reported in the literature for non-accreting objects, see the example of the transition disc around J160421.7 with a $\sim$70\,AU hole and non accreting based on the H$\alpha$ EW but no upper limit on the mass accretion rate \cite{2016A&A...585A..58V}. In addition, disc photoevaporation models do not currently include grain growth and migration, which does occur in the $\sim$1-10\,Myr period over which discs evolve and disperse (see Section~\ref{sect:obs}a for observables). This means that the fraction of predicted optically thick dust discs is in fact an upper limit, thus reducing the discrepancy between theory and observations for the Stage~3 discs.

The population of discs that is most difficult to explain, not only via photoevaporation models, is that of discs that have large dust cavities ($>$20AU) and are accreting vigorously, at rates not too dissimilar from those of full discs ($\sim 10^{-8}$M$_\odot$/yr
\cite{2014A&A...568A..18M}, and Figure~\ref{fig:MaccRhole}).
This population is dominated by millimeter-bright, i.e. massive, dust discs with many of them around early-type stars \cite{2012MNRAS.426L..96O}. These are unlikely to be short-lived objects caught in the act of dispersing their discs 
\cite{2016PASA...33....5O}.
Their nature is hotly debated in the literature. 
It has been shown that 3-6 giant planets are needed to open gas gaps with the observed widths (15-45\,AU) and depletion (factors of 10-1,000) combined with a very low, rather unusual, disc viscosity ($<$0.001) \cite{2016ApJ...825...77D} . In addition, reproducing the $\sim 10$\% transition disc fraction would require extending the incidence of giant planets within $\sim 2$\,AU out to 3-30\,AU, which seems unlikely based on current direct imaging statistics 
\cite{2016PASP..128j2001B}. An alternative explanation requires a dead zone, combined with an MHD wind to remove the inner gas in order  explain the gap sizes and depletions  \cite{2016A&A...596A..81P}. However, predicted mass accretion rates are too low in comparison with observed values, a shortcoming also encountered by photoevaporation models. The failure of most models in reproducing the accretion rates boils down to the fact that an $\alpha$-type accretion is assumed in the inner disc, meaning that large gas surface densities are required at the low velocities implied by typical $\alpha$ values.  
It has recently been proposed that gas accreting at transsonic speeds may help solving this problem \cite{2017ApJ...835...59W}. In this scenario very low gas surface densities (e.g. they use a midplane density of $n_H \sim 10^8 cm^{-3}$) are required in the inner disc, which may be consistent with recent observational constraints 
\cite{2016A&A...585A..58V}. The argument is based on the fact that accretion driven by MHD winds may indeed produce the required accretion speeds if the low density gas in the inner disc has a level of ionisation such that it is sufficiently coupled to the magnetic field, but not so much as to drag the magnetic flux inwards. However, they also point out that this picture cannot be applied to full discs or in the outer regions of discs with cavities as the implied gas surface densities are too low in comparison to those inferred observationally via millimeter observations.

\section{Consequences for planet formation}
\label{sect:implications}

The implications and consequences of disc  dispersal on the formation and evolution of planetary systems have been discussed in a recent review published in the Protostar and Planets VI (PPVI) conference proceedings \cite{2014prpl.conf..475A}. In what follows we aim at providing an update, covering work that has appeared since this review.  For the sake of consistency we structure the discussion  according to the (a) chemical effects and impact on planet formation and (b) dynamical effects. 

\subsection{Chemical effects and impact on planet formation}

One aspect of the influence of photoevaporation on the formation of planetesimals which was considered in the PPVI review is the selective removal of virtually dust free material. Grains in the wind launching region will be lifted in the wind according to the force balance between the drag force 
\cite{2005ApJ...627..286T}, gravity and the centrifugal force. Small ($\sim \mu$m size) grains are expected to be entrained in the wind 
\cite{2011MNRAS.411.1104O,2016MNRAS.463.2725H,2016MNRAS.461..742H}, but the majority of the solid mass  will remain in the disc. This effect is further amplified by considering that due to settling 
\cite{2005A&A...434..971D} the wind launching region will only be populated by sub-$\mu$m grains which can be brought up to the base of the wind via turbulence \cite{1995Icar..114..237D}. Photoevaporation is thus expected to increase the dust to gas ratio in the disc, which is an important factor in triggering planetesimal formation via the streaming and gravitational instabilities 
\cite{2010AREPS..38..493C}. There have been studies investigating the role played by internal EUV + external FUV photoevaporation \cite{2005ApJ...623L.149T} and internal EUV-only photoevaporation \cite{2007MNRAS.375..500A} on the formation of planet(esimal)s. Both studies concluded that while planetesimal formation could be triggered in both cases, it would be at a later time in the disc evolution, when the gas surface density is too low to allow the formation of giant planets. While recent observations of ringed structures in young protoplanetary discs suggest that the planet formation process may have already begun, well before photoevaporation becomes dominant,  this process remains relevant for the formation of terrestrial planets or debris disc  \cite{2008ARA&A..46..339W}. 
 
Elements of a 1D dust evolution model  \cite{2011A&A...525A..11B} have been bootstrapped onto a 1+1D chemical model to investigate the interplay between dust evolution and disc dispersal \cite{2015ApJ...804...29G}. This lead to the conclusion that dust evolution hardly affects the disc lifetimes, while on the other hand large reductions ($\sim$ 2-10) in the gas/dust mass ratio are achieved due to photoevaporation, suggesting that this process may make conditions favorable for the formation of planetesimals by instabilities.

\begin{figure}[b]
\centering
\includegraphics[width=0.75\textwidth]{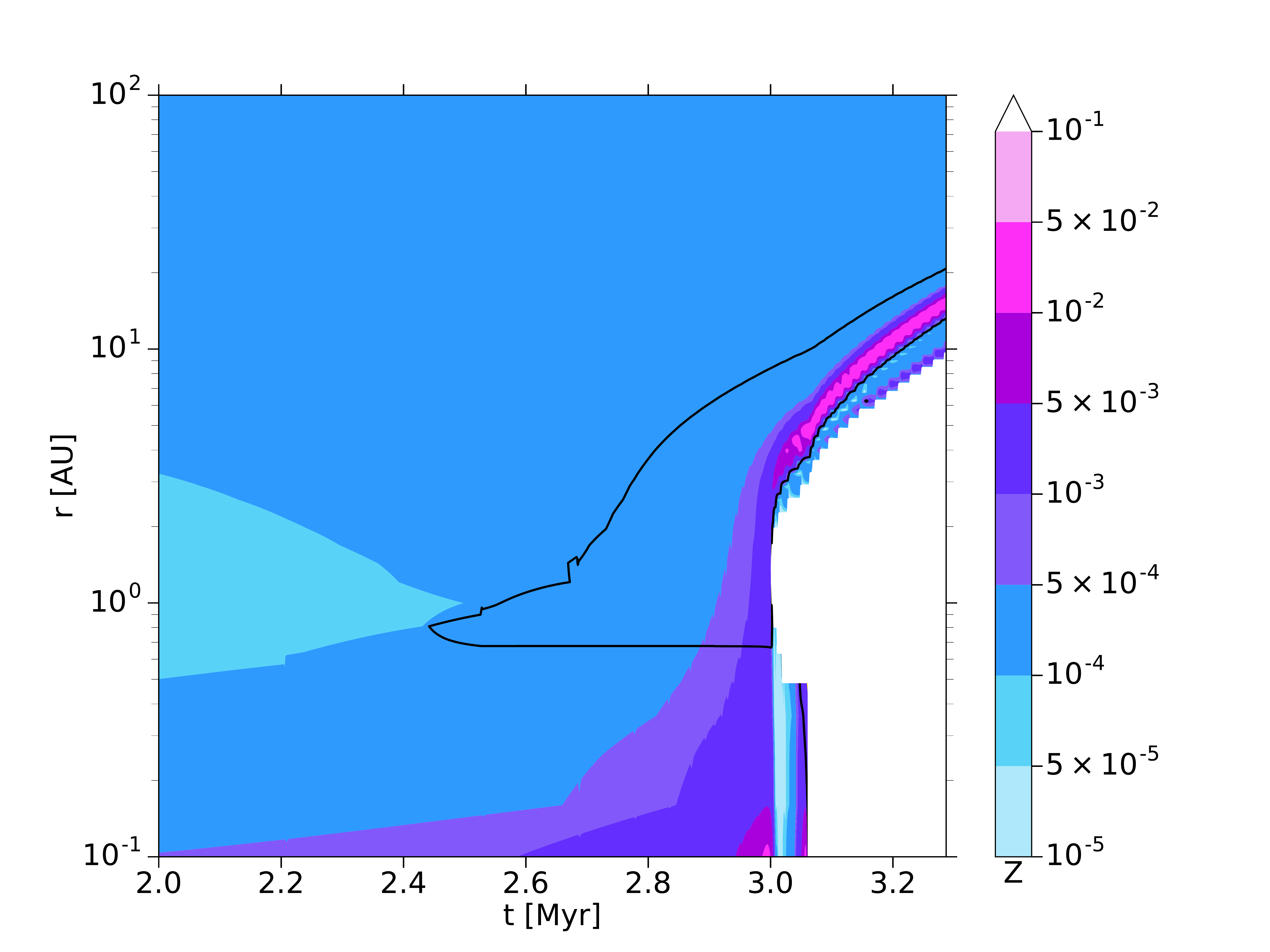}
\caption{The colour map shows the vertically integrated dust-to-gas ratio in the disc at times just before and after the opening of the gap by X-ray photoevaporation. The overlain  black contours shows the location where the Stoke's number are between values of 0.01 and 0.1. 
}\label{fig:si}
\end{figure}

It should be noted that dust to gas ratio is not the only necessary ingredient to allow  convergence of radial drift to form particle clouds leading to the streaming instability. The size of the particles (or better the Stokes number) also plays an important role. It has been shown that for metallicities (column integrated dust-to-gas ratios) greater than $\sim 10^{-2}$ Stokes numbers of order $10^{-2}$ are sufficient to trigger the instability  \cite{2014A&A...572A..78D}. More recent work  extend the condition down to Stoke numbers of order $10^{-3}$ for similar metallicities \cite{2015A&A...579A..43C}. 
Very recent calculations of dust evolution and radial drift in a viscously evolving disc affected by X-ray photoevaporation (Ercolano, Jennings \& Rosotti 2017, in prep), suggest that these criteria may easily be met for standard disc and dust models \cite{2010MNRAS.401.1415O, 2012A&A...539A.148B}. 
Figure~\ref{fig:si} shows a  colour map of the dust to gas ratio with overlain a black contour showing the location where the Stoke's numbers are between values of 0.01 and 0.1.
In this model the low dust to gas mass ratio at 2Myr ($\sim$10$^{-4}$ instead of the canonical $\sim$10$^{-2}$) is due to radial drift which has removed the large particles (which contain most of the dust mass) from the disc. The gap opens  shortly after 3 My and creates a region of pressure maximum inside the gas-rich inner edge of the outer disc where the conditions may become favourable for triggering the streaming instability. This region moves out as the hole grows, possibly leaving behind a population of planetesimals which may provide the seeds  for a terrestrial planetary system in tight configuration, reminiscent of the compact multi-transiting planet systems discovered by Kepler (but note that a similar configuration can also be obtained through  different physical processes \cite{2014ApJ...780...53C}). More detailed models and more thorough investigation of the parameter space are now needed before  more robust conclusions can be drawn. The larger mass loss rates and the more extended photoevaporation profile, characteristic of an X-ray driven wind, solve the problem encountered by previous calculations where, without 
an efficient particle trapping mechanism 
\cite{2012A&A...538A.114P}, radial drift depleted the outer region of an evolving disk before photoevaporation
could operate \cite{2005ApJ...627..286T, 2012MNRAS.423..389H}.

Further effects of internal EUV and external FUV photoevaporation on  the chemical composition of the material in the disc available for planet formation have also been explored  \cite{2006MNRAS.367L..47G, 2015ApJ...798....9M}. These works show that the preferential removal of H/He rich gas by photoevaporation could result in  the gradual enrichment of refractory elements in the disc midplane. This process could explain the larger than expected Ar, Kr, and Xe abundances measured in the atmosphere of Jupiter by the Galileo probe \cite{1999Natur.402..269O}. The models presented to date do not include internal X-ray  and FUV photoevaporation, which, due to their larger rates, may yield  significant enrichment without the need to invoke an external FUV component.  In a very recent study models have been presented for the formation of the solar system incorporating disc evolution, pebbles and gas accretion, type I and II migration, simplified disc photoevaporation and solar system chemical measurements \cite{2017MNRAS.464.4282A}. The models show that photoevaporation of the protoplanetary disc was needed to explain the formation of Jupiter and Saturn with all their constraints.  These findings motivate future work to incorporate more realistic photoevaporation prescription in models for the formation of our Solar System. 

\subsection{Dynamical effects on the formation and evolution of planetary systems}

Trapping dust particles by vortices in discs is an attractive avenue to create "hotspots" for planet formation. 
The possibility of vortices created at the gap edges in transitional discs, by planet formation or photoevaporation has recently been investigated \cite{2014ApJ...795...53Z, 2016MNRAS.458.3927B}. In theory both scenarios involve a pile up of gas at the gap edge which could be then subject to the Rossby-wave instability, leading to the formation of a vortex. 

The removal of gas from the protoplanetary disc  spells the end for gas giant formation, and it provides a natural stop to migration processes involving exchange of angular momentum between the planets and the gas in the disc. The general consequences of  gas removal on the formation and evolution of planetary systems have been explored in a number of seminal works reviewed during PPVI \cite{2014prpl.conf..475A}.  In the last few years progress has been made on the modeling of  giant planet migration within a photoevaporating disc \cite{2015MNRAS.454.2173R}, and it has been shown that the details of the photoevaporation profile, i.e. the driving radiation, as well as  magnitude of the mass loss rates may leave detectable signature on the observed semi-major axis distribution of giant planets. That is to say that photoevaporation does not leave a generally predictable signature, as the effects depend on the profile and the total mass loss rates \cite{2015MNRAS.450.3008E}. These models expand on previous work which explored the effects of EUV photoevaporation on planetary systems architectures \cite{2012MNRAS.422L..82A,2009ApJ...704..989A} by including disc dispersal by X-ray photoevaporation. Recent simulation work by \cite{2016MNRAS.460.2779C} aims at understanding the diversity of observed planetary systems, in particular of the two observed giant planet populations: hot and cold Jupiters. They conclude that a radially structured protoplanetary disc, due to the combination of magnetospheric cavities and photoevaporative winds is needed to at least qualitatively explain the observations. 
The predictive power of these models remains, however, limited due to uncertainties  implicit  in the approach, specifically  on  the rate of mass and angular momentum accretion across the planet's gap 
\cite{2006ApJ...641..526L}.

Moving onto later times of dynamical evolution,  recent work has shown that the disc dispersal mechanism can influence migration histories of planeatary embryos, finally determining the formation of hot super-Earths versus giant planet cores \cite{2014A&A...569A..56C}. This results from resonant chains being disrupted by late dynamical instabilities triggered by the dispersal of the gaseous disc.  The simulations can reproduce many of the exoplanets trends, including the observed giant exoplanet – stellar metallicity correlation and the lack of such correlation for super-Earths \cite{2004A&A...415.1153S,2012Natur.486..375B,2016AJ....152..187M}.

More recent works also highlight the importance of disc dispersal in understanding the migration histories of planetary embrios \cite{2016ApJ...832...83H,2016ApJ...822...90M}.

\section{Summary and Concluding Remarks}\label{sect:summary}
We have summarized empirical constraints on the evolution and dispersal of planet-forming discs and discussed theoretical models that offer a physical interpretation. Particular emphasis was placed on the subset of discs that may be caught in the act of dispersing, often called transition discs.
As it transpires from this review, observations and theory have advanced significantly in just the past few years and are challenging classical views of disc evolution and dispersal. By confronting theory with observations we can draw the following main conclusions: 

\begin{itemize}
\item Current data are consistent with accretion being the main disc dispersal mechanism on $\sim$Myr timescales but cannot prove whether disc viscosity or MHD winds drive accretion.
\item The relatively long disc lifetime (a few Myr) but short disc dispersal timescale ($\sim 10^5$\,yr) can be explained by: i) photoevaporation taking over long-lived viscously evolving discs; ii) MHD winds if the inner disc can retain most of its magnetic flux while the outer disc loses it during evolution; or iii) a combination of the two, where a thermally driven wind, threaded by magnetic lines, exerts a torque on
the gas remaining in the disk and drives accretion.
\item Disc winds are now directly detected toward many young stars. Winds arising inside the EUV gravitational radius for 10,000\,K gas are most likely MHD driven while those tracing material further out may be MHD or photoevaporative in origin
\item The SED-identified transition discs are an heterogeneous group of systems. About 50\% of them can be explained by models that include viscous evolution and photoevaporation. The other 50\%, those that have large mass accretion rates ($\sim 10^{-8}$M$_\odot$/yr) and disc masses, might not be truly dispersing discs and remain difficult to account for in the classical $\alpha$ disc paradigm.
\item Gas diagnostics are starting to reveal cavities and gaps at radial distances comparable to dust cavities in transition discs providing valuable constraints on the ways discs evolve and disperse.
\end{itemize}

At this stage, clarifying the role of accretion and disc winds in the evolution and dispersal of planet-forming discs is critical to make progress in the field and connect the first stages of planet formation to mature planetary systems.
Measurements of turbulence in discs will be key to assess the role of viscous accretion. Luckily, they are becoming possible thanks to the sensitivity and spatial resolution of ALMA and multiple approaches are being tested to understand uncertainties and biases 
\cite{2016MNRAS.460.2779C}. Expanding the set of wind diagnostics to molecular tracers will be also invaluable to clarify the origin of disc winds beyond the EUV gravitational radius as only FUV-driven photoevaporative winds can be molecular and these are launched beyond $\sim$2\,AU \cite{2009ApJ...705.1237G}. Finally, extending the measurements of magnetic fields to disk sizes and mapping their evolution in time seems necessary to test if MHD winds can truly drive accretion and disperse discs from inside-out as observed. As such measurements remain challenging even with current facilities, in the short term empirical constraints may come from mapping the evolution of MHD and photoevaporative winds traced via gas emission lines. 

From the theory side, there are also several important directions that need to be pursued. As discussed in Section~3b, photoevaporative models are mature but still incomplete. A longer-term but challenging goal would be to build a model that include all stellar heating sources, treat disc chemistry and dust evolution, and include 
hydrodynamics to properly determine wind rates and profiles as well as to predict emission line profiles. Global MHD disc simulations, with enough vertical extent, are still needed to demonstrate that MHD winds can drive accretion and disc dispersal. Predictions that could distinguish this mode of evolution from viscously evolving photoevaporating discs are also necessary to make progress in the field. The relative extent of the gas and dust and gas depletions are some of the diagnostics that are becoming  accessible to observers and might be promising to establish how discs evolve and disperse.

Finally, the last few  years have seen several attempts at quantitatively account for disc dispersal in planet formation and evolution models. What is apparent is that disc dispersal is not only a boundary condition to planet formation models, but many of the processes involved can play a powerful role in the assembly of material to make planetesimals, as well as in the evolution of young planetary systems.  Thus the era of quantitative planet formation modeling can only truly begin once a more quantitative picture of disc dispersal is established via comprehensive and observationally constrained theoretical modeling. 

\section*{Data Availability}
The datasets supporting this article have been uploaded as part of the supplementary material.

\section*{Competing Interests}
We have no competing interests.

\section*{Author's Contributions}
B.E. contribution focused on the description of theory, while IP contribution focused on the description of the observations. All authors gave final approval for publication. 

\section*{Funding}
B.E. is funded by a professorship at the Ludwig-Maximilians University of Munich, Germany. I. P. acknowledges support from an NSF Astronomy \& Astrophysics Research Grant (ID: 1312962). The review also benefited from collaborations and/or information exchange within NASA Nexus for Exoplanet System Science (NExSS) research coordination network sponsored by NASA Science Mission Directorate.

\section*{Acknowledgement}
We thank Cornelis Dullemond and a second anonymous referee for their comments which helped improve the review. We also thank J. Owen, G. Picogna, and G. Rosotti for providing transition discs model tracks and synthetic line profiles and S. Edwards, U. Gorti, and N. van der Marel for useful discussions. 

\bibliographystyle{vancouver}
\bibliography{references,bib_Sects1_2_Table}

\end{document}